\documentclass[twocolumn,showpacs,assymb,amsmath]{revtex4}
\usepackage{mathrsfs}
\usepackage{txfonts}
\usepackage{graphicx}
\usepackage{amssymb}
\usepackage{amsmath}
\usepackage{color}
\newcommand{\Tr}{\text{Tr}}

\usepackage{color}
\definecolor{MyDarkBlue}{rgb}{0,0.08,0.45}
\definecolor{yellow}{rgb}{0.99,0.99,0.70}
\definecolor{white}{rgb}{1.0,1.0,1.0}
\definecolor{black}{rgb}{0.00,0.00,0.00}
\definecolor{green}{rgb}{0.8,0.98,0.83}
\pagecolor{green}

\begin{document}
\title[Short Title]{Enhanced exciton transmission by quantum-jump-based feedback}
\author{Y. Q. Ji,$^{1,2}$, M. Qin$^{1,3}$, X. Q. Shao$^{1,2}$, and X. X. Yi$^{1,2}$\footnote{E-mail:
yixx@nenu.edu.cn}}
\address{$^1$Center for Quantum Sciences and School of Physics, Northeast Normal University, Changchun, 130024, People's Republic of China\\
$^2$Center for Advanced Optoelectronic Functional Materials Research, and Key Laboratory for UV Light-Emitting Materials and Technology
of Ministry of Education, Northeast Normal University, Changchun 130024, People's Republic of China\\
$^3$School of Physics and Optoelectronic Technology, Dalian University of Technology, Dalian 116024, People's Republic of China}

\begin{abstract}
With rotating-wave approximation (RWA), we show in this paper that exciton transmission in a one-dimensional two-level molecule chain embedded in a cavity can be enhanced or suppressed by strong cavity-chain couplings. This exciton transmission is closely related to the number of molecules and the distribution of molecular exciton energy. In addition, we propose a proposal to enhance the exciton transmission by quantum-jump-based feedback. These results may find applications in experiments of exciton transmission in organic materials.
\end{abstract}
 \maketitle

\section{Introduction}\label{One}
In organic materials, excitons act as the intermediates between light and charge~\cite{0001,01,02,03}. Excitons usually suffer from relatively large propagation losses associated with decoherence during transmission. Therefore, how to make this transmission more efficient over long distances becomes an important issue in a variety of fields, such as photosynthesis and artificial devices. For photosynthesis, exciton is created with the absorption of a photon, and then it is transferred between the pigments of light-harvesting complexes until it arrives at the the photosynthetic reaction center~\cite{0001,01,02,03}. For organic solar cells, Ref.~\cite{04} demonstrated that significantly enhanced exciton diffusion lengths can be realized in SubPc and the power conversion efficiency can be improved by optimizing the intermolecular separation.

Recently, many works focus on exciton transmission in a molecule chain~\cite{05,06,07,071}, in particular the authors of Ref.~\cite{05} show that if the molecules are strongly coupled to a cavity mode, the exciton conductance can be enhanced by several orders of magnitude. Optical cavities with organic molecules have been extensively studied~\cite{08,09,10,11,12,13}, as it provides us with a new method to study strong coupling effects. In several fields of physics, features caused by strong coupling has been observed. Light-matter interaction can enter into the strong coupling regime by exchanging photons faster than any competing dissipation processes. This is normally achieved by placing the material in a confined electromagnetic fields---cavity modes. The strongly interaction between molecule and a resonant cavity mode leads to the formation of two hybridized light-matter polaritons states, which are separated by the Rabi splitting~\cite{131,132,133,134,135,136,137}.

However, most physical process will be destroyed by the influence of external environments. In order to avoid or decrease the influence of the external environment, many methods have been presented, such as structured environment~\cite{016,017}, decoherence free subspace~\cite{018,019}, quantum Zeno control~\cite{020}, and dynamical control~\cite{021,022}. It is well known that quantum feedback control is a promising method~\cite{023,0231,shao,Carvalho,Hou,Stevenson} to prolong coherence in quantum systems. For instance, one of the simplest systems involving the atom-cavity interaction is the coupling of a two-level atom inside a single-mode cavity, this system is dissipative and the coherence is fast lost, by the feedback control we can prolong the coherence and increase the steady-state entanglement~\cite{024,025}. Based on the continuous monitoring, together with an appropriate choice for the feedback Hamiltonian, the entanglement generation can be well realized ~\cite{026,027}.

Inspired by the result of previous studies~\cite{05,06,07,023,024,025,026,027}, we investigate the influence of molecular number and molecular exciton energy distribution on the exciton conductance. We find that strong coupling could enhance or suppress exciton conductance depending on the molecular number when
molecular exciton energy is identical to each other. Whereas when the molecular exciton energy is randomly distributed, the suppression effect of strong coupling will be affected. In addition, the enhancement of exciton transmission by a quantum-jump-based feedback scheme is also proposed and discussed in this paper.

The remainder of the paper is organized as follows: in Section \ref{Two}, we introduce the basic model for exciton transmission. In Section \ref{Three}, we investigate influence of molecular number and molecular exciton energy distribution on exciton transmission without quantum-jump-based feedback. In Section \ref{Four}, we propose a proposal to enhance the exciton transmission by quantum-jump-based feedback in strong coupling regime. In Section \ref{Five}, we conclude the results.

\section{The model}\label{Two}
\begin{figure}
\centering\scalebox{0.35}{\includegraphics{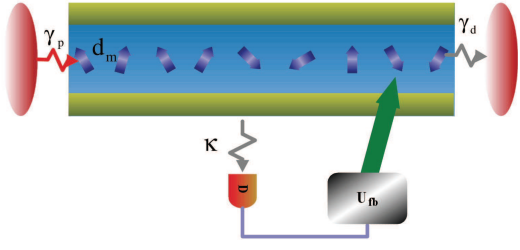}}
\caption{\label{P1}(color online). Schematic illustration for exciton transmission in one-dimensional chain of two-level molecules.
The first molecule is pumped by the left reservoir and the pump rate is $\gamma_{p}$, and the exciton current can be measured on the last molecule on right-hand side with decay rate $\gamma_{d}$. Conditioned on the measurement of the output of the leaky cavity, a Hamiltonian is applied to one of the molecules by a feedback.}
\end{figure}

The model~\cite{05} consists of a chain of $M$ two-level molecules that are embedded in a cavity as shown in Fig.~\ref{P1}.
The coupling strength between the molecules and the single mode cavity is $g_{m}$. The dipole-dipole coupling between the molecules are also considered. Such a system can be described by the following Hamiltonian,
\begin{eqnarray}\label{02}
H=\omega_{0}a^{\dagger}a&+&\sum_{m}\omega_{m}\sigma_{m}^{+}\sigma_{m}^{-}+\sum_{m}g_{m}(a^{\dagger}\sigma_{m}^{-}+\sigma_{m}^{+}a)\cr
&+&\sum_{m,n}V_{mn}(\sigma_{m}^{+}\sigma_{n}^{-}+\sigma_{n}^{+}\sigma_{m}^{-}),
\end{eqnarray}
where $\omega_{0}$ and $\omega_{m}$, are cavity mode energy and the $m$-th molecular exciton energy, respectively, $a$ ($a^{\dagger}$) corresponds to the annihilation (creation) operator of cavity mode (electric field $\vec{E}_{c}(\vec{r}_{m})$), $\sigma_{m}^{+}$ and $\sigma_{m}^{-}$ are Pauli operators on the $m$-th molecule, the coupling strength between cavity and the $m$-th molecule is $g_{m}=-\vec{d}_{m}\cdot\vec{E}_{c}(\vec{r}_{m})$.
If the interaction is strong enough to overcome decoherence effects, two new eigenstates of the system will be formed, separated by what is known as the Rabi splitting energy. So the total coupling between the cavity and molecules can be further characterized by the collective Rabi frequency $\Omega=2\sqrt{\sum_{m}|g_{m}|^{2}}$.
Strong coupling discussed in this paper means $\Omega>|\gamma-\kappa|/2$ with zero detuning $\omega_{0}=\omega_{m}$ as used in~\cite{Savona}. Dipole-dipole interaction (in the quasistatic limit) can be described by
\begin{eqnarray}\label{03}
V_{mn}=\frac{\vec{d}_{m}\cdot\vec{d}_{n}-
3(\vec{d}_{m}\cdot\hat{R}_{mn})(\vec{d}_{n}\cdot\hat{R}_{mn})}{4\pi\epsilon_{0}R_{mn}^{3}},
\end{eqnarray}
with $\hat{R}_{mn}=(\vec{r}_{m}-\vec{r}_{n})/R_{mn}$ and $R_{mn}=|\vec{r}_{m}-\vec{r}_{n}|$. $\vec{d}_{m}$  is the dipole moment and $\vec{r}_{m(n)}$
is the position of the $m(n)$-th molecule.

In order to investigate exciton transmission, we first determine the steady state of the system when the first molecule is incoherently pumped. This system has only one driving term and the state of system is described by its density operator $\rho$. The master equation of the whole system can be expressed in the Lindblad form
\begin{eqnarray}\label{04}
\dot{\rho}=&-&i[H,\rho]+\sum_{m=1}^{M}\gamma_{d}L_{\sigma_{m}^{-}}[\rho]
+\sum_{m=1}^{M}\gamma_{\phi}L_{\sigma_{m}^{+}\sigma_{m}^{-}}[\rho]\cr
&+&\kappa L_{a}[\rho] +\gamma_{p}L_{\sigma_{1}^{+}}[\rho],
\end{eqnarray}
where the terms $L_{s}[\rho]\equiv s\rho s^{\dag}-\frac{1}{2}\{s^{\dag}s,\rho\}$ contain all incoherent process considered here. $\kappa$ denotes the decay rate of the cavity, $\gamma_{d}$ is the decay rate of the molecule including nonradiative and radiative ones, and $\gamma_{\phi}$ is pure dephasing rate. The pumping rate $\gamma_{p}$ is considered to be small such that the system stays in the linear regime. The exciton conductance can be calculated via~\cite{05}
\begin{eqnarray}\label{05}
\sigma_{e}=\frac{\gamma_{d}\Tr(HL_{\sigma_{M}^{-}}[\rho_{ss}])}{\gamma_{p}}.
\end{eqnarray}

\section{Exciton transmission without quantum-jump-based feedback}\label{Three}
The Hamiltonian (\ref{02}) is widely adopted to characterize one-dimensional chain of molecules. We consider the zero-excitation and single-excitation subspace, and choose the molecule parameters as follows~\cite{25,26,27}: $\omega_{0}=2.11$ eV, radiative decay rate $\gamma_{r}=1.32\cdot10^{-6}$ eV, nonradiative decay rate $\gamma_{nr}=1.10$ meV, and  $\gamma_{d}=\gamma_{r}+\gamma_{nr}$, $\gamma_{\phi}=26.3$ meV accounts for pure dephasing. The cavity decay rate is given by $\kappa=0.1$ eV. Dipole moment can be determined by the molecule parameters $\gamma_{r}=\omega_{m}^{3}d^{2}/(3\pi\epsilon_{0}\hbar c^{3})$, giving $d\approx 36$ D (Debye units,$1~\rm{D}=3.33564\times10^{-30}~\rm{C}\cdot\rm{m}$). In addition, the strong dipole-dipole interaction leads to an additional small energy shift $\Delta$ for regular distribution, and the zero detuning corresponds to $\omega_{0}=\omega_{m}+\Delta$.

\subsection{Influence of molecular number on exciton transmission}
\begin{figure}
\centering\scalebox{0.40}{\includegraphics{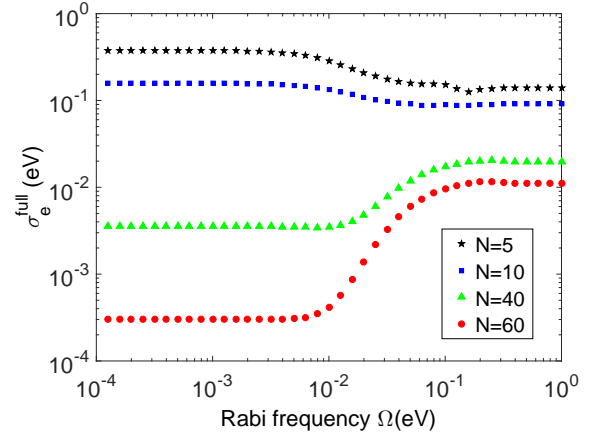}}
\caption{\label{P2}(color online). Exciton conductance as a function of $\Omega$ for full model at zero detuning which molecular number $N=5$, $N=10$, $N=40$ and $N=60$. Strong coupling shows the enhanced role for exciton conductance when the molecular number is large, while it shows suppression effect when   the molecular
number is relatively small.}
\end{figure}
In Fig.~\ref{P2}, we show the exciton conduction $\sigma_{e}$ as a function of the collective Rabi frequency $\Omega$ for different numbers of molecules, at zero cavity-molecule detuning $\omega_{0}=\omega_{m}$. We observe that, for large number of molecules, $\sigma_{e}$ increases with $\Omega$, i.e. the coupling strength between the molecules and the cavity mode can enhance the transmission. When the number of molecules is relatively small, $\sigma_{e}$ decreases with $\Omega$. This result is not exactly the same as that in Ref.~\cite{05}. We now explore the mechanism underlying this observation.

\begin{figure}
\begin{minipage}[ht]{0.9\linewidth}
\centering\includegraphics[width=\textwidth]{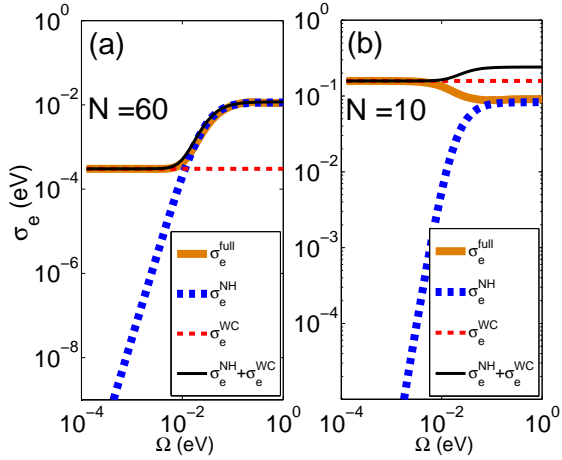}
\end{minipage}
\caption{\label{P3}(color online). Exciton conductance as a function of Rabi frequency $\Omega$ at zero detuning which molecular number $N=10$(a)
and $N=60$(b). The result  is the  same as that in  ref~\cite{05} with $N=60$, but it is different for $N=10$. The solid orange curves is the  full conductance. The dashed red line is the result in the weak coupling limit, i.e.,  $\Omega=0$. The dashed blue
line is the exciton conductance without hopping i.e., the dipole-dipole interactions is turned off. The solid black curves
is a simple  sum over  the dashed red line and dashed blue line.}
\end{figure}
In this model, there are two channels for exciton transmission: one is that in the weak coupling limit ($\Omega=0$), in this case the exciton transmission only relies on the hopping rate between the molecules, and we will name this channel A for simplicity, the exciton conductance is $\sigma_{e}^{WC}$; The other is that in the strong coupling limit, the conductance depends on the cavity-mediated contribution without hopping, we denotes this channel B and the exciton conductance is $\sigma_{e}^{NH}$ in this limit. {In short, the $\sigma_{e}^{WC}$ ($WC$ denotes weak coupling between the cavity and molecule) and $\sigma_{e}^{NH}$ ($NH$ denotes no hopping between the molecules) are the conductances in the weak coupling limit and without hopping, respectively.} The whole exciton conductance is denoted by $\sigma_{e}^{full}$. Starting from Eq.~(\ref{02}), we calculate  numerically the influence of molecular number on the exciton transmission. We find that strong coupling enhances or suppresses exciton conductance in this model. Fig.~\ref{P3} shows the exciton conductance as a function of $\Omega$ at zero detuning $\omega_{0}=\omega_{c}$. When $N$ is large, such as $N=60$, we recover the result obtained in Ref.~\cite{05}, the exciton conductance (solid black curves) is a sum of dashed red line($\sigma_{e}^{WC}$) and dashed blue line($\sigma_{e}^{NH}$) and it
approximately equals to solid orange curve ($\sigma_{e}^{full}$) as shown in Fig.~\ref{P3}(a). However, the result is different to Ref.~\cite{05} if $N$ is relatively small. From the Fig.~\ref{P3}(b), we can see that $\sigma_{e}^{full}\neq\sigma_{e}^{WC}+\sigma_{e}^{NH}$.
{In a nutshell, this is because exciton transmission takes the dominant channel, this does not mean that the other channel is completely absent. In the weak coupling limit, the dominant channel is A, and in the strong coupling limit, the dominant channel is B. Thus, in the strong coupling limit, it will lead to $\rm {min}\left\{\sigma_{e}^{NH},\sigma_{e}^{WC}\right\}<\sigma_{e}^{full}<\rm {max}\left\{\sigma_{e}^{NH},\sigma_{e}^{WC}\right\}$ and $\sigma_{e}^{full}$ closer to $\sigma_{e}^{NH}$.}

\begin{figure}
\centering\scalebox{0.45}{\includegraphics{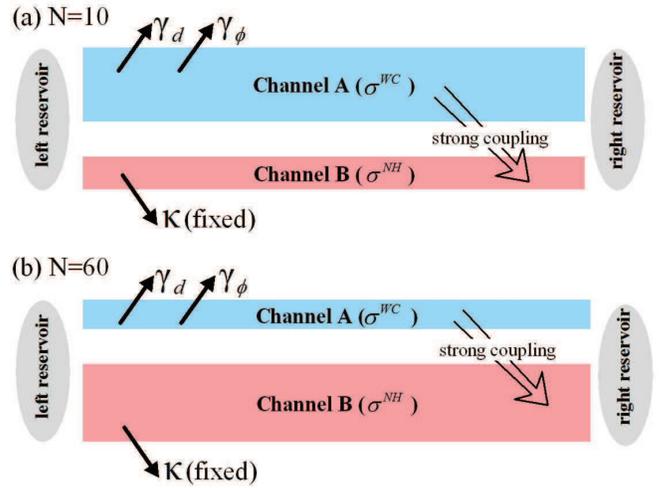}}
\caption{\label{P4}(color online). Schematic illustration. Two transmission channels A and B, the width of line (A and B) represents the size of transmission efficiency. In the case of strong coupling, channel B is the dominated transmission channel. Black arrow with ``strong coupling" denotes the change of dominated channel from A to B when the coupling strength increases. (a)For $N=10$, $\sigma^{WC}>\sigma^{NH}$, i.e., strong coupling suppresses the whole exciton conductance. (b)For $N=60$, $\sigma^{WC}<\sigma^{NH}$, i.e., strong coupling enhances the whole exciton conductance.}
\end{figure}
From Eq.~(\ref{04}), we know that, with fixed cavity-decay rate $\kappa$ and pumping rate $\gamma_{p}$, more molecules leads to more decay through the incoherent processes described by terms with $\gamma_{d}$ and $\gamma_{\phi}$. Namely, for less molecules (e.g. $N=10$), exciton transmission through channel A is more efficient than that through channel B, while for more molecules (e.g. $N=60$), it is exactly the opposite, see Fig.~\ref{P4}.
With fixed $\kappa$ and $\gamma_{p}$, the efficiency through channel B is a constant. Meanwhile, the increasing of the coupling strength between the molecules and the cavity mode, i.e., the collective Rabi frequency $\Omega$, would result in the change of dominant transmission channel from  A to B. In other words, for any number of molecules, channel A dominates the exciton transmission in the weak coupling limit, while channel B does in the strong coupling limit. Thus, if the efficiency of channel B is greater than the efficiency of channel A, strong coupling enhances the whole exciton conductance $\sigma_{e}^{full}$. Otherwise, strong coupling suppresses $\sigma_{e}^{full}$. In Fig.~\ref{P4}, channel B ($\sigma_{e}^{NH}$) denotes the channel which dominates  in the strong coupling limit, where the exciton can efficiently bypass the chain of molecules. And in the weak coupling limit, exciton transmission through channel B ($\sigma_{e}^{NH}$) can be ignored.

\begin{figure}
\centering\scalebox{0.4}{\includegraphics{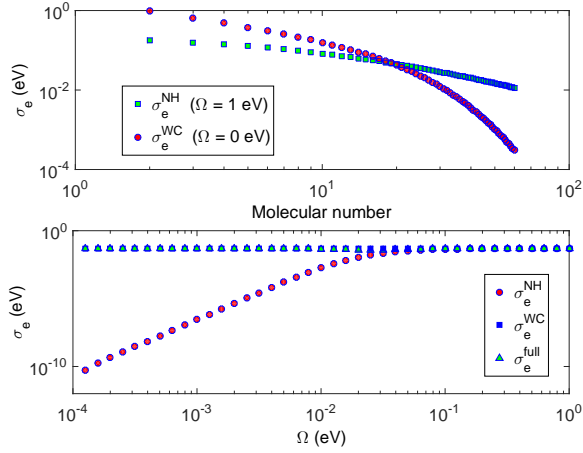}}
\caption{\label{P5}(color online). (a) The exciton conductance $\sigma_{e}^{NH}$ and $\sigma_{e}^{WC}$ as a function of molecular number.
(b) $\sigma_{e}^{NH}$ and $\sigma_{e}^{WC}$ as a function of $\Omega$ when $N=19$, $\sigma_{e}^{NH}=\sigma_{e}^{WC}$ in strong coupling regime.}
\end{figure}
These discussions naturally raise a question: with the parameters we adopt for this model, when does the exciton transport through channel A as efficiently
as that through channel B? Or when are the exciton conductance $\sigma_{e}^{NH}=\sigma_{e}^{WC}$? In Fig.~\ref{P5}(a), we plot the exciton conductance
$\sigma_{e}^{NH}$ and $\sigma_{e}^{WC}$ as a function of molecular number. It is observed that $N=19$ acts as the critical value at which $\sigma_{e}^{NH}=\sigma_{e}^{WC}$. In the case of less molecules, strong coupling suppresses the exciton conductance $\sigma_{e}^{full}$ (see Fig.~\ref{P5}(a)) due
to $\sigma_{e}^{NH}<\sigma_{e}^{WC}$. On the contrary, for more molecules $\sigma_{e}^{NH}>\sigma_{e}^{WC}$ when the coupling is strong. In Fig.~\ref{P5}(b) we plot the exciton conductance as a function of $\Omega$ for $N=19$. Strong coupling neither shows the enhanced
or suppression effect due to $\sigma_{e}^{NH}=\sigma_{e}^{WC}$.

\subsection{Influence of molecular exciton energy on exciton transmission}
\begin{figure*}
\begin{minipage}[ht]{0.49\linewidth}
\centering\includegraphics[width=\textwidth]{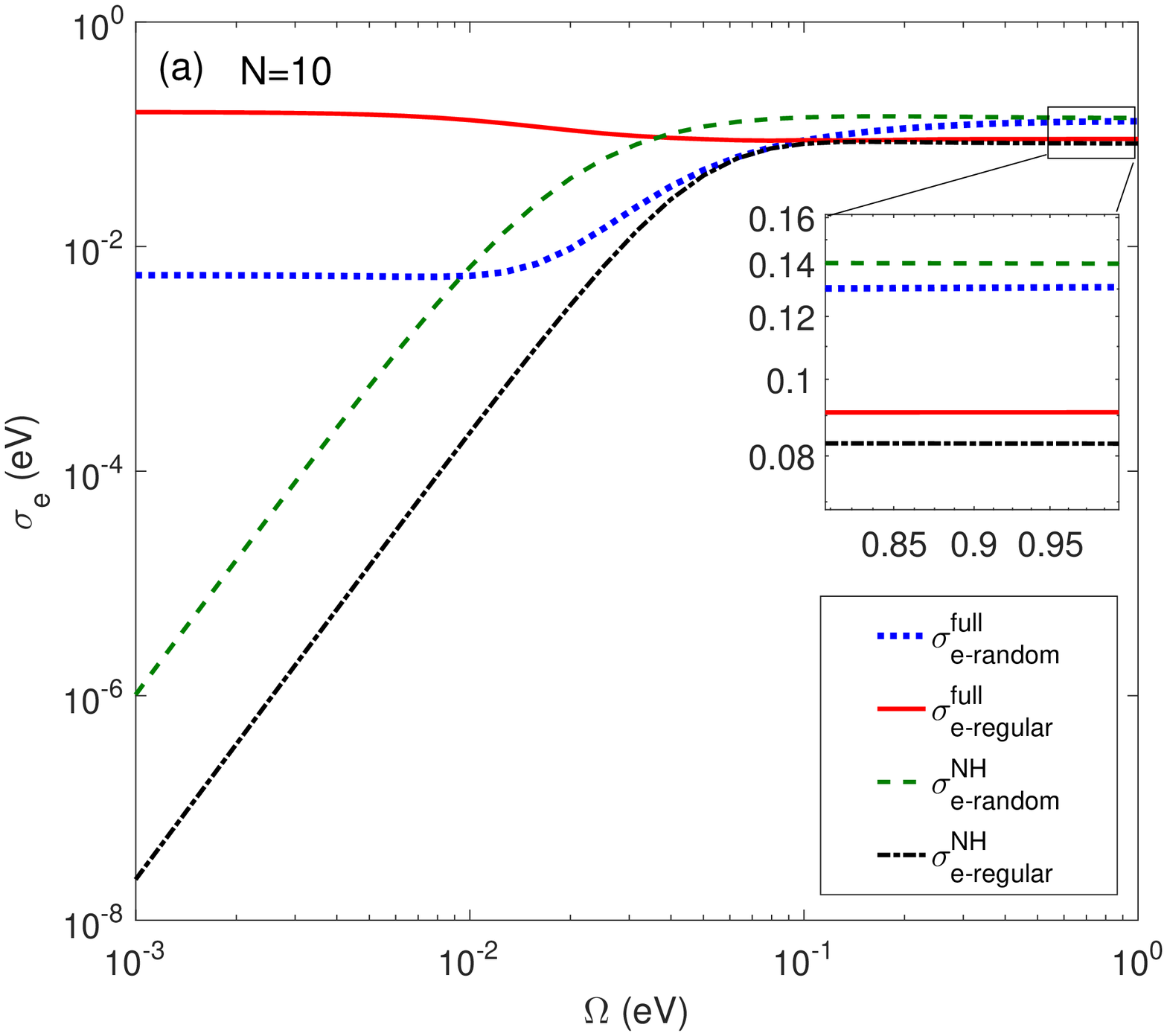}
\end{minipage}
\hfill
\begin{minipage}[ht]{0.49\linewidth}
\centering\includegraphics[width=\textwidth]{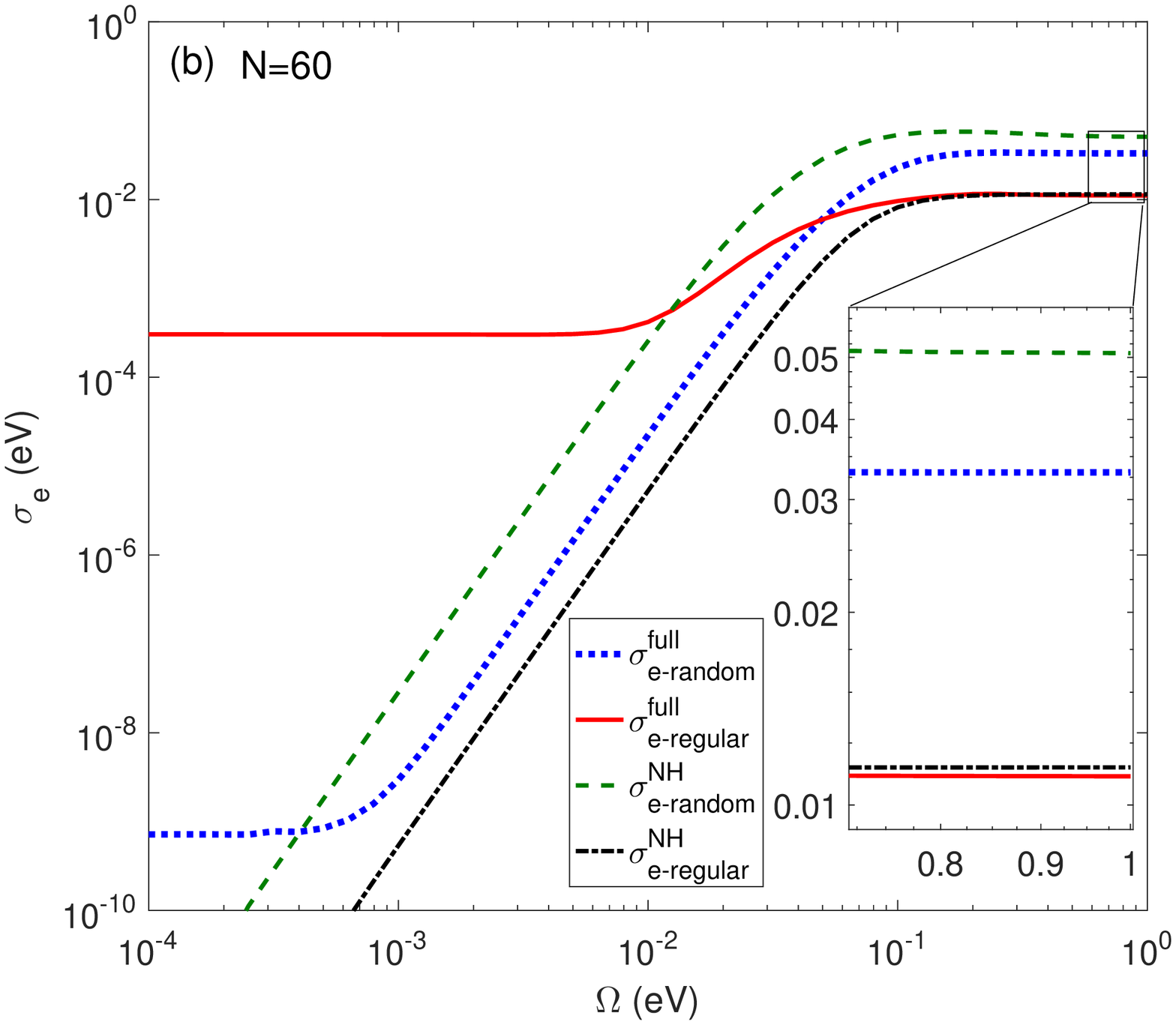}
\end{minipage}
\caption{\label{P6}(color online). Exciton conductance as a function of $\Omega$ when the molecular exciton energy $\omega_{m}$ takes a random normal distribution, where the mean is $p=2.11$ eV with standard deviation $q$. The curves are obtained by average over many numerical simulations. (a) and (b) shows the influence of the molecular exciton energy on the exciton transmission. $N=10$ and $N=60$ with standard deviation 0 and 0.211 eV, respectively.}
\end{figure*}

In this section, we analyse the influence of random distribution of the molecular exciton energy on the exciton transmission. For $N=10$ in the weak coupling limit, the exciton transmission only depends on the dipole-dipole interaction of the molecules, the random distribution of molecular exciton energy suppresses exciton transmission. Therefore, the exciton conductance is far less than the case of equal exciton energy. The greater the variance $q$ is, the larger the suppression, and the smaller the exciton conductance $\sigma_{e}^{WC}$. While in the strong coupling limit, channel B dominates the exciton transmission, in this case it main depends on cavity-mediated strength rather than the hopping rate of the molecules, the exciton transmission is enhanced greatly. From the other point of view, $\sigma_{e}^{NH}$ and $\sigma_{e}^{WC}$ determines whether strong coupling could enhance the exciton transmission or not. This means $\sigma_{e}^{NH}>\sigma_{e}^{WC}$ ensures that strong coupling could enhance exciton transmission when $\omega_{m}$ is randomly distributed, otherwise strong coupling suppresses exciton transmission. This feature does not appear for more molecules(for example $N=60$), since $\sigma_{e}^{NH}>\sigma_{e}^{WC}$ always satisfies. As shown in Fig.~\ref{P6}, we plot the exciton conductance as a function of $\Omega$ when $\omega_{m}$ [$\omega_{m}\sim N(p,q^{2})$] takes a normal distribution with the mean $p$ and standard deviation $q$. The curves are given by an average over many  numerical simulations. $\omega_{m}$ is the molecular exciton energy and the mean is $p=2.11$ eV with standard deviation $q$. If the $q$ is too small, it will not change $\sigma_{e}^{WC}$ obviously. Hence, the mean $p$ and standard deviation $q$ are chosen as 2.11 eV and 0.211 eV, respectively, to obtain a apparent effect.

{In addition, the Fig.~\ref{P6} appears that randomness can increase the conductance, this is because when the exciton energy of each molecule is equal, it is resonant with the cavity. However, when we keep the first and the last molecular exciton energy constant and other ones normally distributed, the interaction between the molecules and the cavity is effectively suppressed, it leads the exciton to bypass the intermediate molecules and transport from the first molecular to the last one more effectively. Thus, the transmission efficiency can be enhanced greatly.}

\section{enhanced exciton transmission by quantum-jump-based feedback}\label{Four}
In order to increase exciton transmission, we introduce a feedback scheme based on the continuous monitoring of quantum jumps together with the feedback Hamiltonian. As we will show the feedback can lead to an improvement of exciton transmission. Note that the feedback is conditioned on the measurement result and thus must act after the measurement. This can be done by introducing a description of the measurement scheme and feedback. The schematic diagram is shown in Fig.~\ref{P1}, the cavity dissipation, i.e. $L_{a}[\rho]$, is measured by a photodetector D whose signal provides the input in the control Hamiltonian $F$.

In this kind of monitoring, the absence of signal predominates and the control is triggered only after a detection click, i.e. a quantum jump occurs. The unconditioned master equation for this model was derived by Wiseman in~\cite{023} and, for our system, it reads
\begin{eqnarray}\label{06}
\dot{\rho}=&-&i[H,\rho]+\sum_{m}\gamma_{d}L_{\sigma_{m}^{-}}[\rho]+\sum_{m}\gamma_{\phi}L_{\sigma_{m}^{+}\sigma_{m}^{-}}[\rho]\cr
&+&\kappa L_{U_{fb}a}[\rho] +\gamma_{p}L_{\sigma_{1}^{+}}[\rho].
\end{eqnarray}
It is easily to explain jump feedback with $L_{U_{fb}a}[\rho]$ in Eq.~(\ref{06}): $L_{U_{fb}a}[\rho]=U_{fb}a\rho a^{\dag}U_{fb}^{\dag}-\frac{1}{2}\{a^{\dag}a,\rho\}$.
The unitary transformation $U_{fb}=exp(-iF\delta t/\hbar)$, representing the finite amount of evolution imposed by the control Hamiltonian on the system, is chosen to be applied according to the detection result, which is described by the action of $a$ in the first term of the superoperator $L_{a}[\rho]$.

\begin{figure*}
\begin{minipage}[ht]{0.49\linewidth}
\centering\includegraphics[width=\textwidth]{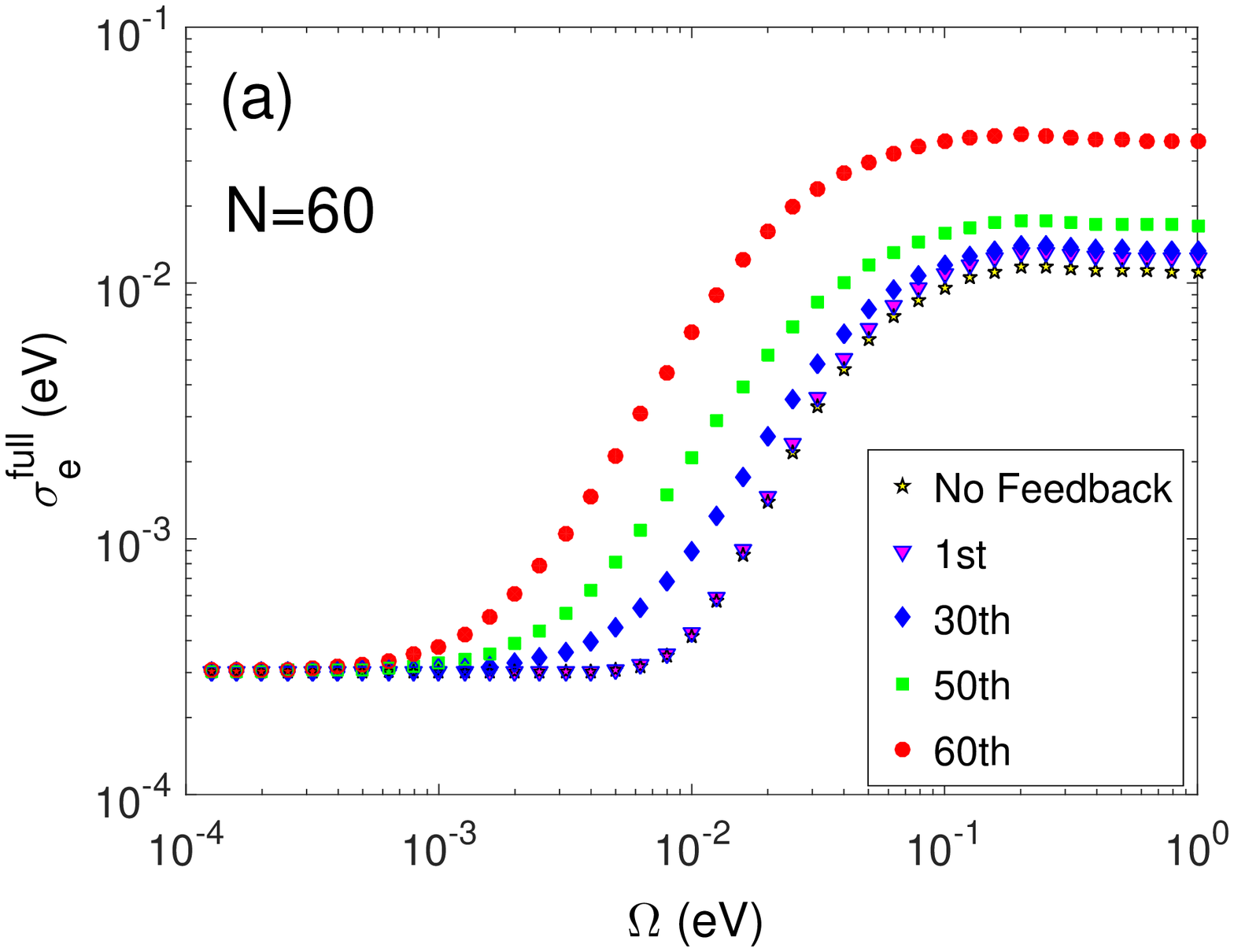}
\end{minipage}
\hfill
\begin{minipage}[ht]{0.49\linewidth}
\centering\includegraphics[width=\textwidth]{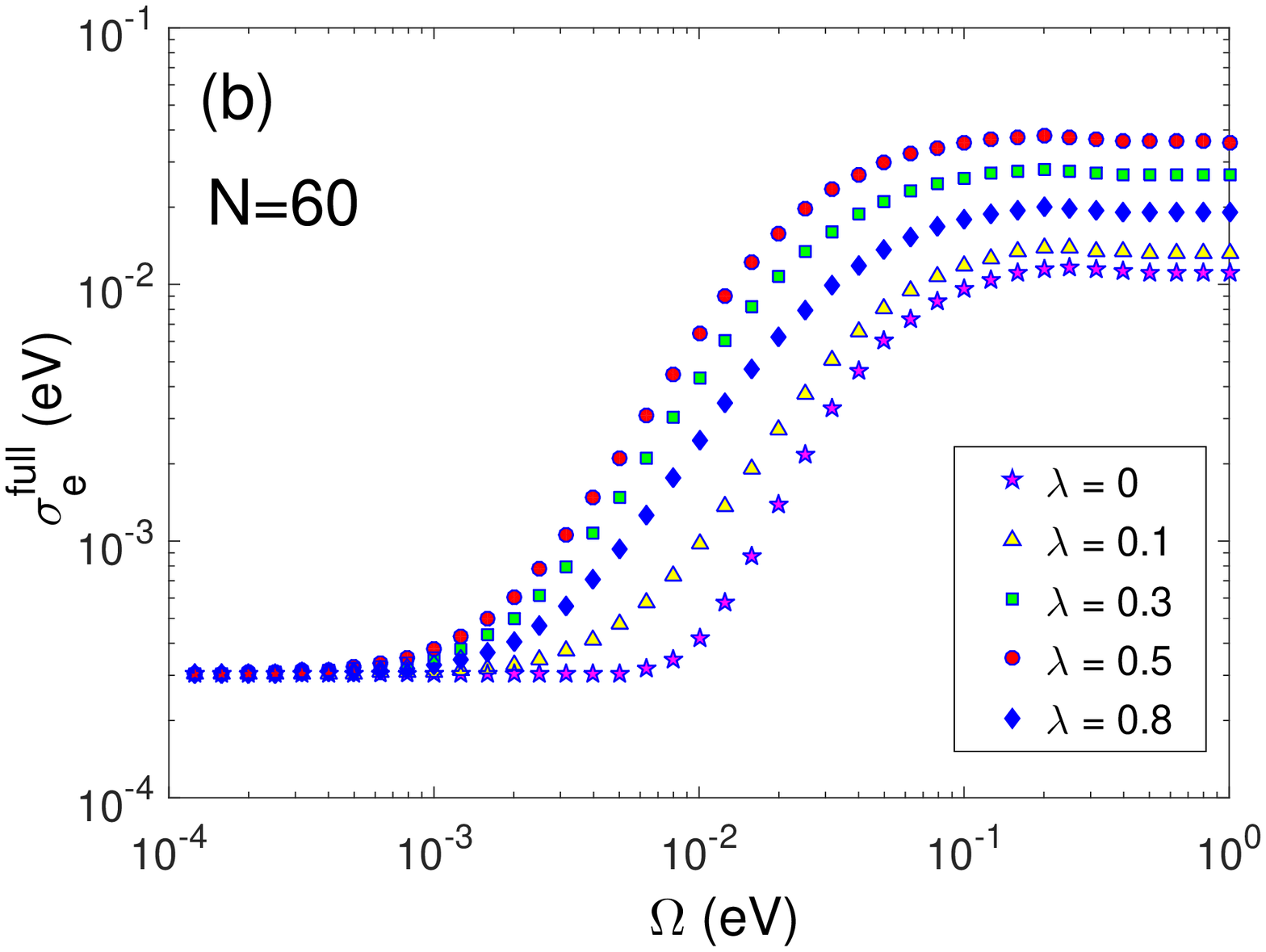}
\end{minipage}
\hfill
\begin{minipage}[ht]{0.49\linewidth}
\centering\includegraphics[width=\textwidth]{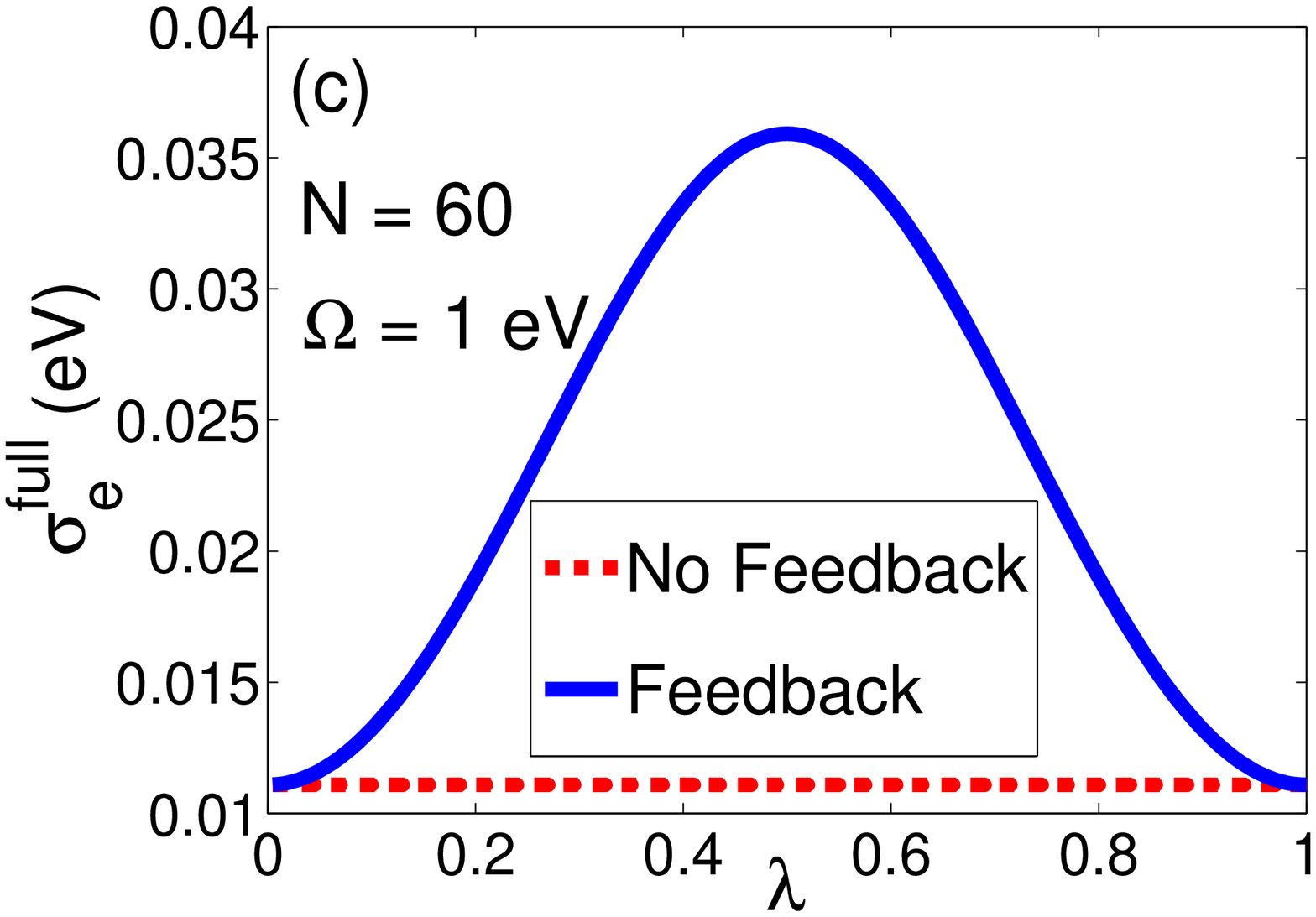}
\end{minipage}
\hfill
\begin{minipage}[ht]{0.49\linewidth}
\centering\includegraphics[width=\textwidth]{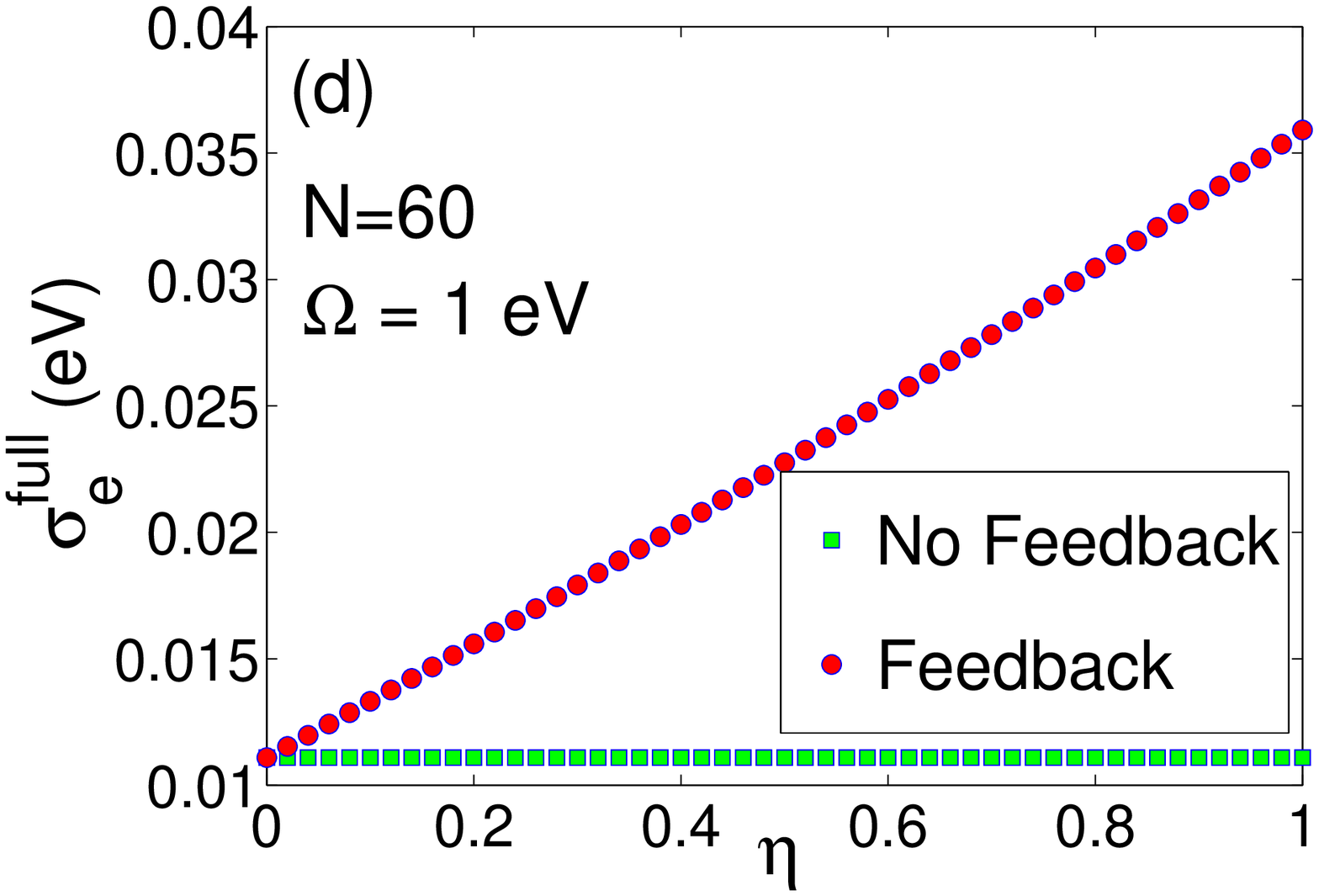}
\end{minipage}
\caption{\label{P7}(color online). The exciton conductance as a function of $\Omega$ when the control acts on 1st, 30th, 50th, 60th molecule, respectively, for N=60 and $\lambda = 0.5$ is shown in (a). (b)For different $\lambda$, the exciton conductance changes with $\Omega$. (c)The exciton conductance as a function of $\lambda$ and reach maximum when $N=60$, $\Omega=1eV$ and $\lambda=0.5$. (d)The exciton conductance as a function of detector efficiency $\eta$ when $N=60$, $\Omega=1$ eV and $\lambda=0.5$.}
\end{figure*}

Starting from Eq.~(\ref{06}), we choose the local feedback $U_{fb}=exp(-i\tilde{\lambda}\sigma_{N}^{x})$ with $\tilde{\lambda}=\lambda\pi$ and $\sigma_{N}^{x}=\sigma_{N}^{+}+\sigma_{N}^{-}$, which is simple and realizable in experiment~\cite{28}, where the control acts on just one of the molecules. In the jump-based feedback, the corresponding exciton conductance is shown in Fig.~\ref{P7}(a) as a function of $\Omega$ and the feedback parameters. The feedback is acted on 1st, 30th, 50th, 60th molecule, respectively, only after a detection clicks. The exciton conductance is enhanced obviously when the control is applied on the last molecule in the strong coupling regime, {i.e. the more closer to the end of the chain, the more conductance increases.} However, the exciton conductance has no significant change in the  weak coupling regime since it only depends on the dipole-dipole interaction of molecules but not on the cavity-medium.
In addition, we find that the exciton conductance depends on the feedback amplitude $\lambda$ as shown in Fig.~\ref{P7}(b). Fig.~\ref{P7}(c) is the plot of exciton conductance versus driving amplitude, it shows that the maximum locates at $\lambda=0.5$. In this case the maximum exciton conductance is about 0.036 eV, with appropriate choice of the feedback amplitude $\lambda=0.5$ and $N=60$ in the strong coupling regime. Nevertheless, exciton conductance is about 0.011 eV if the feedback is cut off.

In Eq.~(\ref{06}), the effects of detection efficiency has been neglected in the detection process, which is important for this system, given that feedback relies on the manipulation of the system based on information gained by the measurement result. It has two distinct situations when a jump occurs: firstly, the detector clicks and the feedback $U_{fb}$ is applied on molecule; secondly, the detector fails to click and there is no control action. We can use detection efficiency $\eta$ to describe this circumstance:
\begin{eqnarray}\label{07}
\dot{\rho}=&-&i[H,\rho]+\sum_{m}\gamma_{d}L_{\sigma_{m}^{-}}[\rho]+\sum_{m}\gamma_{\phi}L_{\sigma_{m}^{+}\sigma_{m}^{-}}[\rho]\cr
&+&\kappa \eta L_{U_{fb}a}[\rho] +\kappa (1-\eta) L_{a}[\rho] + \gamma_{p}L_{\sigma_{1}^{+}}[\rho].
\end{eqnarray}
When the detector efficiency $\eta$ is zero, no information is extracted from the measurement result and the master equation reduces to Eq.~(\ref{04}) where the feedback is cut off. Obviously, in the limit of perfect detection, Eq.~(\ref{06}) is recovered, and, for a local feedback, exciton transmission can be further enhanced even if the detector efficiency is not perfect. We have plotted the exciton conductance changes with detector efficiency $\eta$ when $N=60$ and $\Omega=1$ eV in Fig.~\ref{P7}(d), the  maximum and minimum are consistent with the above analysis. In short, the feedback can enhance exciton transmission.

\begin{figure}
\begin{minipage}[ht]{0.95\linewidth}
\centering\includegraphics[width=\textwidth]{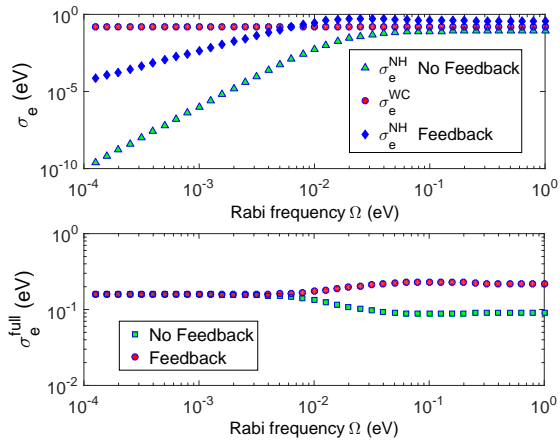}
\end{minipage}
\caption{\label{P8}(color online). (a)$\sigma_{e}^{NH}$ as a function of $\Omega$ when the molecular number $N=10$ with
feedback is cut off and turn on, respectively, feedback makes $\sigma_{e}^{NH}>\sigma_{e}^{WC}$ in strong coupling regime.
(b)$\sigma_{e}^{full}$ as a function of $\Omega$ and the feedback can change suppression effect of strong coupling .}
\end{figure}

We have concluded that the feedback can enhance the exciton conductance with large number of molecules. As to the case with small number of molecules, we plot the exciton conductance $\sigma_{e}^{full}$ as a function of $\Omega$  with and without the feedback for $N=10$. The feedback increases  $\sigma_{e}^{NH}$ and leads to $\sigma_{e}^{NH}>\sigma_{e}^{WC}$ in the strong coupling regime, as shown in Fig.~\ref{P8}(a). We also find that the suppression  effect of strong coupling for small molecules is destroyed. In this case, strong coupling does not suppress $\sigma_{e}^{full}$, as shown is Fig.~\ref{P8}(b). Hence, the feedback can change suppression effect of strong coupling effectively when the molecular number is small.

\section{conclusion}\label{Five}
In conclusion, we have investigated the influence of molecular number and molecular exciton energy distribution on the exciton transmission through strong coupling in systems composed of a one-dimensional chain of molecule inside a cavity under RWA. When the coupling is strong enough (weak enough), exciton transmission only depends on cavity-mediated contribution (the hopping of molecules). Strong coupling can be beneficial or detrimental to exciton transmission depending on the molecular number, the transmission is also connected with whether molecular exciton energy is identical or randomly distributed. We have also discussed the effect of quantum-jump-based feedback and detector efficiency on the exciton transmission.

\section*{ACKNOWLEDGMENTS}
This work is supported by National Natural Science Foundation of China (NSFC) under Grants No. 11534002, No. 61475033
and Fundamental Research Funds for the Central Universities under Grants No. 2412016KJ004.

\end{document}